\documentclass[11pt]{article}
\font\cero=cmbx10 scaled 1728 %
\font\uno=cmcsc10 scaled 1200 %
\font\dos=cmti10 scaled 1200 %
\font\tres=cmbx12 scaled 1200 %
\title{\cero Hamiltonians and Lagrangians of
non-autonomous one-dimensional mechanical systems}
\author{{\uno G.F.\ Torres del Castillo} \\
{\dos Departamento de F\'{\i}sica Matem\'atica, Instituto de Ciencias} \\
{\dos Universidad Aut\'onoma de Puebla, 72570 Puebla, Pue., M\'exico} \\
{\uno I. Rubalcava Garc\'{\i}a} \\
{\dos Facultad de Ciencias F\'{\i}sico Matem\'aticas} \\
{\dos Universidad Aut\'onoma de Puebla, Apartado postal 1152,} \\
{\dos 72001 Puebla, Pue., M\'exico}}
\date{ }
\begin{document}
\maketitle
\section*{ }
It is shown that a given non-autonomous system of two first-order
ordinary differential equations can be expressed in Hamiltonian
form. The derivation presented here allow us to obtain previously
known results such as the infinite number of Hamiltonians in the
autonomous case and the Helmholtz condition for the existence
of a Lagrangian.\\[1ex]
{\it Keywords:} Non-autonomous systems; Hamilton equations;
Lagrangians \\[2ex]
Se muestra que un sistema dado, no aut\'onomo, de ecuaciones
diferenciales ordinarias de primer orden puede expresarse en forma
hamiltoniana. La deducci\'on presentada aqu\'{\i} nos permite
obtener resultados previamente conocidos tales como el n\'umero
infinito de hamiltonianas en el caso aut\'onomo y la
condici\'on de Helmholtz para la existencia de una lagrangiana.\\[1ex]
{\it Descriptores:} Sistemas no aut\'onomos; ecuaciones de Hamilton;
lagrangianas \\[2ex]
PACS: 45.05.+x; 45.20.-d

\section*{\tres 1. Introduction}
As is well known, it is very convenient to express a given system of
ordinary differential equations (not necessarily related to
classical mechanics) as the Euler--Lagrange equations associated
with some Lagrangian, $L$, or as the Hamilton equations associated
with some Hamiltonian, $H$ (see, {\em e.g.}, Ref.\ 1). One of the
advantages of such identifications is the possibility of finding
constants of motion, which are related to symmetries of $L$ or $H$.
Also, the Hamiltonian of a classical system is usually regarded as
an essential element to find a quantum version of the mechanical
system.

In the simple case of a mechanical system with forces derivable from
a potential (that may depend on the velocities), there is a
straightforward procedure to find a Lagrangian or a Hamiltonian.
However, in the case of non-conservative mechanical systems or of
systems not related to mechanics, the problem of finding a
Lagrangian or a Hamiltonian is more involved. A given system of $n$
second-order ordinary differential equations are equivalent to the
Euler--Lagrange equations for some Lagrangian if and only if a set
of conditions (known as the Helmholtz conditions) are fulfilled
(see, {\em e.g.}, Refs.\ 2, 3 and the references cited therein).

The aim of this paper is to give a straightforward procedure to find
a Hamiltonian for a given system of two first-order ordinary
differential equations (which may not be equivalent to a
second-order ordinary differential equation) that possibly involves
the time in an explicit form. The results derived here contain the
Helmholtz condition for $n = 1$ (in the case where the given system
is equivalent to a second-order equation). In Sec.\ 2 the main
results of this paper are established, demonstrating that a given
system of first-order ordinary differential equations can be
expressed in Hamiltonian form looking for an integrating factor of a
differential form made out of the functions contained in the system
and, in Sec.\ 3, several examples are presented. In Sec.\ 4 we show
that, in the appropriate case, our results lead to the Helmholtz
condition for the existence of a Lagrangian.

\section*{\tres 2. Hamiltonians and canonical variables}
We shall consider a system of first-order ordinary differential
equations of the form
\begin{equation}
\dot{x} = f(x,y,t), \qquad \dot{y} = g(x,y,t), \label{1}
\end{equation}
where $f$ and $g$ are two given functions. A system of this class
can be obtained from a second-order equation
\[
\ddot{x} = F(x, \dot{x}, t),
\]
by introducing the variable $y = \dot{x}$. We are initially
interested in finding a Hamiltonian function, $H$, and canonical
variables, $q$, $p$, such that the corresponding Hamilton's
equations be equivalent to the system (\ref{1}).

Assuming that there is an invertible relation between the variables
$x$, $y$ and a set of canonical coordinates $q$, $p$, $x =
x(q,p,t)$, $y = y(q,p,t)$, in such a way that Eqs.\ (\ref{1}) are
equivalent to the Hamilton equations for $q$ and $p$ with a
Hamiltonian $H$, making use of the chain rule, one finds that
\begin{equation}
-g {\rm d} x + f {\rm d} y = \frac{\partial(x,y)}{\partial(q,p)}
{\rm d} H - \frac{\partial y}{\partial t} {\rm d} x + \frac{\partial
x}{\partial t} {\rm d} y + {\rm terms\ proportional\ to\ } {\rm d}
t. \label{0}
\end{equation}
Therefore, given the system (\ref{1}) we start by considering the
differential form
\begin{equation}
- (g - \phi) {\rm d} x + (f - \psi) {\rm d} y, \label{2}
\end{equation}
where
\[
\phi(q,p,t) \equiv \frac{\partial y(q,p,t)}{\partial t}, \qquad
\psi(q,p,t) \equiv \frac{\partial x(q,p,t)}{\partial t}
\]
are functions unspecified by now (see Eq.\ (\ref{7.5}) below). For a
fixed value of $t$, the differential form (\ref{2}) is always
integrable (see any standard text on ordinary differential
equations, {\em e.g.}, Ref.\ 4); that is, there exist (locally)
functions $\sigma$ and $H$, which may depend parametrically on $t$,
such that
\begin{equation}
- (g - \phi) {\rm d} x + (f - \psi) {\rm d} y = \sigma {\rm d} H.
\label{3}
\end{equation}

Now, for simplicity, without any loss of generality (since, once we
have found a set of canonical coordinates, we have the liberty of
making any canonical transformation afterwards), we choose $q \equiv
x$ (hence, $\psi = 0$) and, therefore,
\[
\frac{\partial(x,y)}{\partial(q,p)} = \frac{\partial p}{\partial y}.
\]
Then, by comparing Eqs.\ (\ref{0}) and (\ref{3}), the canonical
momentum, $p$, must be such that
\begin{equation}
\frac{\partial p(x,y,t)}{\partial y} = \frac{1}{\sigma(x,y,t)}.
\label{5}
\end{equation}
Hence
\begin{equation}
{\rm d} p = \frac{\partial p}{\partial x} {\rm d} x +
\frac{1}{\sigma} {\rm d} y + \frac{\partial p}{\partial t} {\rm d} t
\label{6}
\end{equation}
or, equivalently,
\begin{equation}
{\rm d} y = - \sigma \frac{\partial p}{\partial x} {\rm d} x +
\sigma {\rm d} p - \sigma \frac{\partial p}{\partial t} {\rm d} t
\label{6.1}
\end{equation}
thus, recalling that $x = q$, this last expression shows that
\begin{equation}
\phi = - \sigma \frac{\partial p(x,y,t)}{\partial t} \label{6.5}
\end{equation}
and we can also write Eq.\ (\ref{6}) in the form
\begin{equation}
{\rm d} p = \frac{\partial p}{\partial x} {\rm d} x +
\frac{1}{\sigma} {\rm d} y - \frac{\phi}{\sigma} {\rm d} t.
\label{6.6}
\end{equation}
Since this is an exact differential, we have
\begin{equation}
\frac{\partial \sigma^{-1}}{\partial t} = \frac{\partial}{\partial
y} (- \sigma^{-1} \phi) = - \sigma^{-1} \frac{\partial
\phi}{\partial y} - \phi \frac{\partial \sigma^{-1}}{\partial y}.
\label{7.5}
\end{equation}
This equation establishes a relation between the integrating factor
and the function $\phi$ (see examples below).

From Eqs.\ (\ref{3}), with $\psi = 0$, and (\ref{6.6}) we have
\begin{eqnarray*}
{\rm d} H \!\!\! & = & \!\!\! - \frac{1}{\sigma} (g - \phi) {\rm d}
x + \frac{1}{\sigma} f {\rm d} y + \frac{\partial H}{\partial t} {\rm d} t \\
& = & \!\!\! - \frac{1}{\sigma} (g - \phi) {\rm d} x + f \left( {\rm
d} p - \frac{\partial p}{\partial x} {\rm d} x + \frac{\phi}{\sigma}
{\rm d} t \right) + \frac{\partial H}{\partial t} {\rm d} t \\
& = & \!\!\! -  \left( \frac{g}{\sigma} - \frac{\phi}{\sigma} + f
\frac{\partial p}{\partial x} \right) {\rm d} q + f {\rm d} p +
\left( \frac{\partial H}{\partial t} + f \frac{\phi}{\sigma} \right)
{\rm d} t.
\end{eqnarray*}
Hence, considering $H$ as a function of $q$, $p$, and $t$,
\begin{equation}
\frac{\partial H}{\partial p} = f = \dot{q} \label{7}
\end{equation}
[see Eqs.\ (\ref{1})] and
\begin{equation}
- \frac{\partial H}{\partial q} = \frac{g}{\sigma} -
\frac{\phi}{\sigma} + f \frac{\partial p}{\partial x} = \dot{p},
\label{8}
\end{equation}
since, according to Eqs.\ (\ref{6.6}) and (\ref{1}),
\[
\dot{p} = \frac{\partial p}{\partial x} \dot{x} +
\frac{\dot{y}}{\sigma} - \frac{\phi}{\sigma} = \frac{\partial
p}{\partial x} f + \frac{g}{\sigma} - \frac{\phi}{\sigma}.
\]
Equations (\ref{7}) and (\ref{8}) are equivalent to the original
system (\ref{1}) and have the desired Hamiltonian form.

Summarizing, the system of equations (\ref{1}) can be written in the
form of the Hamilton equations, with the Hamiltonian determined by
Eq.\ (\ref{3}) and the canonical momentum defined by Eq.\
(\ref{6.6}).

The fact that the left-hand side of Eq.\ (\ref{3}) multiplied by
$\sigma^{-1}$ is an exact differential yields (when $\psi = 0$)
\[
\frac{\partial}{\partial y} [- \sigma^{-1}(g - \phi)] =
\frac{\partial}{\partial x} (\sigma^{-1}{f}),
\]
which amounts to
\begin{equation}
(g- \phi) \frac{\partial \sigma^{-1}}{\partial y} + \sigma^{-1}
\frac{\partial}{\partial y} (g - \phi) + f \frac{\partial
\sigma^{-1}}{\partial x} + \sigma^{-1} \frac{\partial f}{\partial x}
= 0. \label{7.6}
\end{equation}
Hence, making use of Eqs.\ (\ref{1}), (\ref{7.6}) and (\ref{7.5}),
we obtain
\begin{eqnarray}
\frac{{\rm d}}{{\rm d}t} \sigma^{-1} \!\!\! & = & \!\!\!
\frac{\partial \sigma^{-1}}{\partial x} \dot{x} + \frac{\partial
\sigma^{-1}}{\partial y} \dot{y} + \frac{\partial
\sigma^{-1}}{\partial t} \nonumber \\
& = & \!\!\! f \frac{\partial \sigma^{-1}}{\partial x} + g
\frac{\partial \sigma^{-1}}{\partial y} + \frac{\partial
\sigma^{-1}}{\partial t} \nonumber \\
& = & \!\!\! \phi \frac{\partial \sigma^{-1}}{\partial y} -
\sigma^{-1} \frac{\partial}{\partial y} (g - \phi) - \sigma^{-1}
\frac{\partial f}{\partial x} + \frac{\partial
\sigma^{-1}}{\partial t} \nonumber \\
& = & \!\!\! - \sigma^{-1} \left( \frac{\partial f}{\partial x}
+\frac{\partial g}{\partial y} \right). \label{pH}
\end{eqnarray}
(Note the cancelation of $\phi$.)

Equation (\ref{pH}) shows that the function $\sigma$ is determined
up to a factor that is a constant of motion and, therefore, there
exists an infinite number of Hamiltonians (and, correspondingly, of
expressions for $p$). It may be noticed that Eq.\ (\ref{pH}) is just
Liouville's theorem.

\section*{\tres 3. Examples}
A first example is provided by the equation
\[
\ddot{x} + \gamma \dot{x} + \omega_{0}^{2} x = \eta(t),
\]
where $\gamma$ and $\omega_{0}$ are constants, and $\eta(t)$ is an
arbitrary function, which corresponds to a forced damped harmonic
oscillator. Taking $y = \dot{x}$, we have $\dot{y} = - \gamma y -
\omega_{0}^{2} x + \eta(t)$, which is of the form (\ref{1}) with
$f(x,y,t) = y$, and $g(x,y,t) = - \gamma y - \omega_{0}^{2} x +
\eta(t)$. Then Eq.\ (\ref{pH}) reduces to
\[
\frac{{\rm d}}{{\rm d}t} \sigma^{-1} = \gamma \sigma^{-1}
\]
and we can take $\sigma = {\rm e}^{- \gamma t}$ (any other choice
would require the knowledge of the explicit form of $\eta$) then
from Eq.\ (\ref{7.5}) we see that
\[
\frac{\partial \phi}{\partial y} = - \gamma,
\]
which is satisfied with $\phi = - \gamma y$. Substituting all these
expressions into Eq.\ (\ref{3}) we have (with $t$ treated as a
constant)
\[
\big( \omega_{0}^{2} x - \eta(t) \big) {\rm d} x + y {\rm d} y =
{\rm e}^{- \gamma t} {\rm d} H
\]
and, therefore, we can take $H = {\rm e}^{\gamma t} (y^{2}/2 +
\omega_{0}^{2} x^{2}/2 - \eta(t) x)$. Finally, from Eq.\ (\ref{6.6})
we find that $p$ can be chosen as $p = {\rm e}^{\gamma t} y$. The
corresponding Lagrangian can be calculated in the usual way, by
means of the Legendre transformation.

The results of the previous section allow us to readily derive those
of Ref.\ 5, corresponding to the autonomous case. In fact, when the
functions $f$ and $g$, appearing in Eqs.\ (1), do not depend
explicitly on the time, from Eqs.\ (\ref{3}) and (\ref{1}), taking
$\phi = 0 = \psi$, we have $\sigma \dot{H} = - g \dot{x} + f \dot{y}
= -gf + fg = 0$. This means that $H$ is {\em some}\/ constant of
motion, which is not unique; we can replace it by $H' = G(H)$, with
$G$ being an arbitrary function. $H'$ is also a constant of motion
and $\sigma$ will not depend explicitly on $t$ [see Eq.\
(\ref{7.5})], no matter what (time-independent) Hamiltonian we
choose.

The expressions given above allow us to find $H$, which need not be
related to the total energy. In the example considered in the
appendix of Ref.\ 5, $f(x,y) = y$, $g(x,y) = - ky$, where $k$ is a
constant ({\em i.e.}, $\ddot{x} = - k \dot{x}$). Then, $-g {\rm d} x
+ f {\rm d} y = ky {\rm d} x + y {\rm d} y = y {\rm d}(kx + y)$ and,
therefore, we can take $\sigma = y$ and $H = kx + y$.

We end this section by considering the problem studied in Ref.\ 6
(which corresponds approximately to a relativistic particle
subjected to a constant force, $\lambda$, and a force of friction
proportional to the square of the velocity), namely (with the
appropriate changes in notation)
\[
m \dot{y} = (\lambda - \gamma y^{2}) (1 - \alpha^{2} y^{2}),
\]
where $m$ represents a mass, $\lambda$, $\gamma$, and $\alpha$ are
constants. Thus, $f(x,y) = y$, and $g(x,y) = (\lambda - \gamma
y^{2}) (1 - \alpha^{2} y^{2})/m$. Thus,
\begin{eqnarray*}
-g {\rm d} x + f {\rm d} y \!\!\! & = & \!\!\! - \frac{1}{m}
(\lambda - \gamma y^{2}) (1 - \alpha^{2} y^{2}) {\rm d} x + y {\rm
d} y \\
& = & \!\!\! (\lambda - \gamma y^{2}) (1 - \alpha^{2} y^{2}) \left[
- \frac{{\rm d} x}{m} + \frac{y {\rm d} y}{(\lambda - \gamma y^{2})
(1 - \alpha^{2} y^{2})} \right].
\end{eqnarray*}
Comparing with Eq.\ (\ref{3}) (with $\phi = 0 = \psi$) we
immediately see that we can take
\[
\sigma = (\lambda - \gamma y^{2}) (1 - \alpha^{2} y^{2})
\]
and
\begin{eqnarray*}
H \!\!\! & = & \!\!\! - \frac{x}{m} + \int \frac{y {\rm d}
y}{(\lambda - \gamma y^{2}) (1 - \alpha^{2} y^{2})} \\
& = & \!\!\! - \frac{x}{m} + \frac{1}{2(\lambda \alpha^{2} -
\gamma)} \ln \left| \frac{\lambda - \gamma y^{2}}{1- \alpha^{2}
y^{2}} \right|.
\end{eqnarray*}
According to Eq.\ (\ref{6.6}), the canonical momentum $p$ can be
taken as
\[
p = \int \frac{{\rm d} y}{(\lambda - \gamma y^{2}) (1 - \alpha^{2}
y^{2})}.
\]
Despite the huge difference with the expressions given in Ref.\ 6,
one can show that the Hamiltonian obtained in that reference is
essentially the exponential of our $H$. (See, Eqs.\ (23) and (26) of
Ref.\ 6.)

\section*{\tres 4. The Helmholtz condition}
The case where one starts with a second-order equation of the form
\begin{equation}
\ddot{x} = F(x, \dot{x}, t) \label{11}
\end{equation}
(considered in Refs.\ 2, 3), is a particular case of the treatment
above if one defines, {\em e.g.}, $y \equiv \dot{x}$, that
transforms Eq.\ (\ref{11}) into the system
\[
\dot{x} = y, \qquad \dot{y} = F(x,y,t),
\]
which is of the form (\ref{1}) with $f(x,y,t) = y$ and $g(x,y,t) =
F(x,y,t)$. Then Eq.\ (\ref{pH}) reduces to
\begin{equation}
\frac{{\rm d}}{{\rm d}t} \sigma^{-1} = - \sigma^{-1} \frac{\partial
F}{\partial y},
\end{equation}
which is the Helmholtz condition when there is one degree of freedom
(see, {\em e.g.}, Ref.\ 2 and the references cited therein; note
that $\sigma^{-1} = \partial p/ \partial y = \partial p/
\partial \dot{x} = \partial^{2} L / \partial \dot{x}^{2}$ is the
integrating factor $w_{11}$ employed in these references).

On the other hand, not every system of equations of the form
(\ref{1}) comes from a second-order equation $\ddot{x} = F(x,
\dot{x}, t)$. An example is given by
\[
\dot{x} = f(x,t), \qquad \dot{y} = g(y,t),
\]
where there is no coupling between the variables $x$, $y$. Here
(choosing $\phi = 0 = \psi$)
\[
-g {\rm d} x + f {\rm d} y = fg \left( - \frac{{\rm d} x}{f} +
\frac{{\rm d} y}{g} \right).
\]
Therefore, if we assume that $\sigma = fg$ does not depend
explicitly on of $t$ [see Eq.\ (\ref{7.5})], we can take
\[
H = - \int \frac{{\rm d} x}{f} + \int \frac{{\rm d} y}{g}
\]
and, from Eq.\ (\ref{5}),
\[
p = \int \frac{{\rm d} y}{\sigma} = \frac{1}{f} \int \frac{{\rm d}
y}{g}.
\]
Thus, $H = pf - \int f^{-1} {\rm d} x$ and with the Hamiltonian
being a linear function of $p$, the Legendre transformation is not
defined nor the Lagrangian.

\section*{\tres 5. Concluding remarks}
As we have shown, at least in the case of a system of two
first-order ordinary differential equations, finding a Hamiltonian
is essentially equivalent to finding an integrating factor for a
linear differential form in two variables. The integrating factor
also determines the expression for the canonical momentum. Equation
(\ref{pH}) is analogous to the Helmholtz condition, but, in the
present approach, it leads directly to the Hamiltonian (in the
standard approach, finding a solution to the Helmholtz conditions,
only gives the second partial derivatives $\partial^{2} L/\partial
\dot{x}_{i} \partial \dot{x}_{j}$). When the system is
non-autonomous, it is convenient to find the integrating factor
using Eq.\ (\ref{pH}), while in the autonomous case, it may be more
simply obtained from the linear differential form itself. Finally,
as shown in Sec.\ 4, there are systems of equations for which a
Lagrangian does not exist, but a Hamiltonian description can be
given.

\section*{\tres Acknowledgment}
The authors would like to thank Dr.\ M.\ Montesinos for enlightening
discussions.

\section*{References}
\newcounter{ref} \begin{list}{\hspace{1.3ex}\arabic{ref}.\hfill}
{\usecounter{ref} \setlength{\leftmargin}{2em}
\setlength{\itemsep}{-.98ex}}
\item H.\ Goldstein, {\it Classical Mechanics}, 2nd ed., (Addison-Wesley,
Reading, Mass., 1980).
\item S.A.\ Hojman and L.C.\ Shepley, {\it J.\ Math.\ Phys.}\ {\bf
32} (1991) 142.
\item S.K.\ Soni and M.\ Kumar, {\it Europhys.\ Lett.}\ {\bf 68}
(2004) 501.
\item G.F.\ Simmons, {\it Differential Equations with
Applications and Historical Notes}, 2nd ed., (McGraw-Hill, New York,
1991).
\item G.F.\ Torres del Castillo, {\it Rev.\ Mex.\ F\'{\i}s.}\ {\bf 50} (2004) 379.
\item G.\ Gonz\'alez, {\it Int.\ J.\ Theor.\ Phys.}\ {\bf 43} (2004)
1885.
\end{list}
\end{document}